\documentclass[twocolumn,showpacs,preprintnumbers,amsmath,amssymb]{revtex4}
\usepackage{graphicx}% Include figure files
\usepackage{dcolumn}% Align table columns on decimal point
\usepackage{bm}% bold math

\begin{document}

%\preprint{APS/123-QED}

\title{Observation of elastic collisions between lithium atoms and calcium ions} % Force line breaks with \\

\author{Shinsuke Haze}
\author{Sousuke Hata, Munekazu Fujinaga, Takashi Mukaiyama}%
\affiliation{Institute for Laser Science, University of Electro-Communications, 1-5-1 Chofugaoka, Chofu, Tokyo 182-8585, Japan}
%\affiliation{


\begin{abstract}
 We observed elastic collisions between laser-cooled fermionic lithium atoms and calcium ions at the energy range from 100 mK to 3 K. 
Lithium atoms in an optical-dipole trap were transported to the center of the ion trap using an optical tweezer technique, and a spatial overlap of the atoms and ions was realized in order to observe the atom-ion interactions.
The elastic scattering rate was determined from the decay of atoms due to elastic collisions with ions.
The collision-energy dependence of the elastic scattering cross-section was consistent with semi-classical collision theory. 
\end{abstract}
\maketitle

\section{\label{sec:introduction}Introduction}%:\protect\\ The line
%break was forced \lowercase{via} \textbackslash\textbackslash}
Over the decades, many efforts have focused on observing, understanding, and controlling interactions between particles in the ultracold regime. 
Recent progress in laser cooling technology has made it possible to explore ultracold collisions between neutral atoms and ions, which have so far been inaccessible for rigorous investigation. An ultracold atom-ion hybrid system would also provide a new scheme for the control of neutral atoms with an electric field through atom-ion interactions. Furthermore, the atom-ion hybrid system has many potential applications such as atom-ion sympathetic cooling \cite{Goodman,Ravi,Hall}, ultracold molecular ion production \cite{Hudson}, ultracold superchemistry \cite{Heinzen}, and local probing of quantum degenerated gases \cite{Sherkunov}. 
In the course of this research, there has been pioneering work on efficient charge transfer in homonuclear atom-ion collisions \cite{Grier}, interactions between Bose-Einstein condensates and single ions \cite{Zipkes,Schmid}, and chemical reactions at the single-particle level \cite{Ratschbacher}.
Regarding the local probing of atomic gases, there have been several demonstrations of high-resolution imaging and state detection in an ultracold atomic system \cite{Gericke,Bakr,Gemelke}. 
Since ultracold ions can be manipulated with an electric field that has little influence on neutral atoms, ions will make a useful state-detection head  for neutral atomic gases \cite{Kollath}. 
For the applications mentioned above, it is important to reach ultra-low temperature at which $s$-wave and only a few partial waves contribute to atom-ion collisions. It is also important to understand the collision properties of ultracold atom-ion systems.

In experiments involving ultracold atom-ion hybrid systems, atoms are either trapped in a magneto-optical trap (MOT) \cite{Grier,Goodman,Ravi,Hall} or an optical dipole trap (ODT) \cite{Zipkes,Schmid,Ratschbacher}, while ions are trapped in a radio frequency (RF) trap that has a deep and tight harmonic potential.
The motion of ions in an RF trap is considered to be a combination of secular motion and rapid RF micromotion.
Even though ions are stably trapped in the RF potential, there is still a finite heating due to in-phase motion (micromotion) of ions  to the applied RF field.
Micromotion heating would be a serious issue in precision metrology \cite{Rosenband} or quantum information processing \cite{Blatt} using ions, so efforts have been made to minimize the micromotion heating and suppress the  Doppler shift \cite{Berkeland}.
It is also important to suppress micromotion heating in the atom-ion hybrid system because it limits the lowest accessible temperature, preventing entry into the quantum regime of atom-ion collisions.
Recently, Cetina et al. quantitatively investigated the heating rate arising from the position shift of an ion due to atom-ion interactions, and pointed out that the incoming energy from the driving RF field becomes larger for larger atom/ion mass ratios \cite{Cetina}.
Considering that the energy threshold of an $s$-wave collision is also higher for a combination of light neutral atoms and heavy ions, they concluded that the combination of light atoms and heavy ions is desirable for the study of ultracold atom-ion collisions.
 
Our system contains a combination of $^{40}$Ca$^{+}$ ions and $^{6}$Li atoms that has a smaller atom/ion mass ratio than the atom-ion combinations demonstrated previously.
Therefore, our choice of atom-ion mixture should be less likely to suffer from micromotion heating effects, and has an advantage in creating a suitable mixture for future applications \cite{Cetina}.
Furthermore, as we introduce a fermionic isotope of lithium atoms, a single ion may be used to measure the local thermodynamic quantities of fermionic atoms for the determination of the thermodynamic behavior of strongly-interacting fermionic atoms \cite{Horikoshi, Nascimbene, Ku}.

This article is organized as follows: Section \ref{sec:Experimental setup} describes our experimental setup for the preparation of a mixture consisting of lithium atoms and calcium ions.
In Section \ref{sec:Results and discussion}, we discuss the observation of elastic collisions between atoms and ions, and the elastic scattering cross section is derived from the decay of atoms.
Additionally, we compare the obtained scattering cross section with a theoretical prediction based on a semi-classical approach.
In Section \ref{sec:Conclusion}, we present a summary and outlook of this work.

\section {\label{sec:Experimental setup}Experimental Setup}
\subsection {\label{sec:Atom-Ion mixing apparatus}Atom-Ion mixing apparatus}

\begin{figure}[h]
\includegraphics[width=8cm]{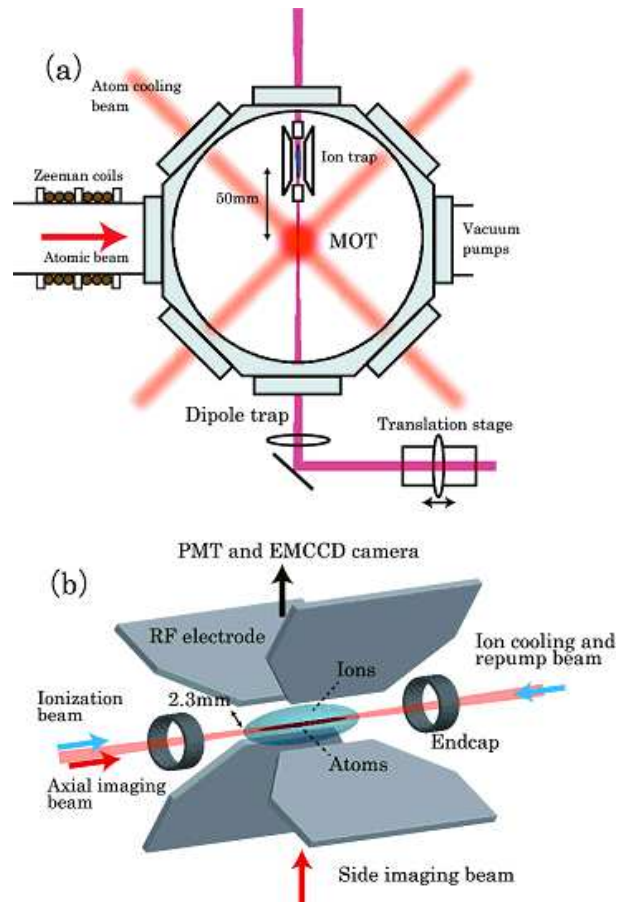}\\
\caption{\label{fig:1} (a) An overview of the experimental setup used to create ultracold atom-ion mixtures. Lithium atoms are trapped in a magneto-optical trap (MOT) and transported to the center of the linear ion trap, which is located 50 mm away from the MOT. (b) A schematic of the linear ion trap composed of four blade and cylindrical end cap electrodes. Atoms in an optical dipole trap and the ion cloud are spatially overlapped at the center of the trap electrodes. } 
\end{figure}

In this section, we present an overview of the experimental setup for the implementation of a mixture of lithium atoms and calcium ions.
A schematic of the apparatus is shown in Fig.\ \ref{fig:1}(a).
Cold atoms and ions are trapped in an ultra-high-vacuum environment evacuated by ion pumps and a titanium sublimation pump.
The resulting pressure is measured to be $P$ $\sim$ 5 $\times$ 10 $^{-11}$ torr. 

An $^{6}$Li atomic beam from an atom oven was decelerated by a Zeeman slower and captured in an MOT at the center of the vacuum chamber.
After compression of the atoms in the MOT, the atoms are loaded into an optical potential formed by a far off-resonant laser light focused at the center of the atomic cloud.
The laser beam for the ODT is aligned along the ion trap axis, which is parallel to a weak confinement axis of ions formed by a DC electric potential.
After the atoms are captured in the ODT, the atoms are transported into the region of the ion trap, which is mounted at a distance of 50 mm from the MOT.
To transport the atoms, the focused position of the ODT laser is moved by translating the position of the focusing lens using a linear air-bearing stage (ABL1500: Aerotech inc.).

The ion trap consists of four blade-shaped RF electrodes and two cylindrical end caps, as shown in Fig.\ \ref{fig:1}(b).
The trapped ions are Doppler-cooled and mixed with lithium atoms for the measurement of collisions between the atoms and the ions.
The position of the RF trap and the laser geometry are carefully designed to avoid intercepting each other. 

Detailed explanations of atom and ion trapping are presented in the following sections.

\subsection {\label{sec:Lithium atoms trapping} Lithium atom trapping}

Lithium atoms are captured in an MOT with red-detuned laser beams incident upon the atoms in a six-beam configuration.
The detuning of the trapping laser is -6 $\Gamma$ from the resonance frequency of the $S_{\rm{1/2}}$ - $P_{\rm{3/2}}$ transition, with $\Gamma$=2$\pi\times$5.8 MHz being the natural linewidth of the relevant transition.
The laser is modulated at a frequency of around 250 MHz (depending on the trapping laser frequency) using an electro-optical modulator to repump from the lower hyperfine states.
For an efficient loading of lithium atoms into an ODT potential, the atoms in the MOT are compressed by increasing the magnetic field gradient and decreasing the cooling laser detuning.
The resulting temperature of the atoms after compression is 280 $\rm{\mu K}$, which was estimated from a time-of-flight (TOF) measurement.
The compression period is 20 ms, and the atoms are pumped to the lowest hyperfine state $\mid F = 1/2, m_{F} = \pm 1/2 \rangle$ at the end of the compression process.
For an ODT of lithium atoms, an infrared (IR) laser beam with a wavelength of 1064 nm is tightly focused onto the center of the atom cloud.
An ODT laser with an intensity of 15 W and a beam size of 38 $\rm{\mu}$m creates a trap depth of 430 $\rm{\mu K}$ at a focused position.
After loading atoms into the ODT, one of the lens optics, which is used for IR beam focusing, is translated with an air-bearing stage by 25 mm, and the corresponding position shift of the focused spot is 50 mm.
The maximum velocity of the stage translation is 20 mm per second.

\begin{figure}
\includegraphics[width=7cm]{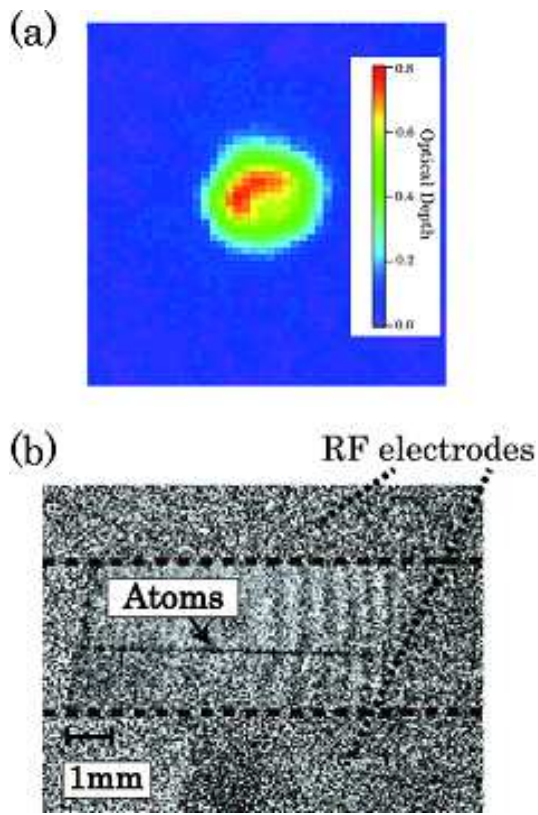}\\
\caption{\label{fig:2} (a) Absorption image of atoms in an ODT. The imaging beam is incident along the direction of  the symmetry axis of the ion trap electrodes. (b) Absorption image of the atoms carried into the ion trap electrode taken in side imaging . }
\end{figure}

Figures\ \ref{fig:2}(a) and (b) show typical absorption images of atoms confined in the ODT after being transported to the middle of the ion trap electrodes. 
Figure\ \ref{fig:2}(a) is an absorption image with atoms  taken along the direction of the ODT laser, which goes through the end cap electrodes (see Fig.\ \ref{fig:1}(b)).
The image was taken just after releasing the atoms from the dipole trap potential.
The number of atoms in the ODT was determined from these axial images because the integration along the symmetry axis of the ODT makes the absorption image robust against the imaging noise.
10$^{4}$ $\sim$ 10$^{5}$ atoms are captured under typical experimental conditions.
The atomic temperature and the radial trapping frequency were 180 $\rm{\mu K}$ and 6.2 kHz, respectively.

A density profile of the atoms along the symmetry axis of the ion trap can also be obtained from absorption imaging by irradiating the resonant light propagating orthogonal to  the ion trap axis (see Fig.\ \ref{fig:1}(b)).
A typical image of the atoms in the side imaging  is shown in Fig.\ \ref{fig:2}(b).
The signal around the center of the electrode shadow indicates  the lithium atoms, and the length of the atomic cloud is $\sim$4 mm (calculated Rayleigh length is 4.0 mm.).
This side imaging scheme was used to check the overlap of the atoms and ions.

\subsection{\label{sec:Calcium ion trap}Calcium ion trap}

$^{40}$Ca$^{+}$ ions were trapped in a linear RF trap, as shown in Fig.\ \ref{fig:1}(b).
The distance between the trap center and the edge of the RF electrodes is 2.3 mm, and the end cap electrodes are separated by 15 mm. 
Aluminum was chosen as the electrode material in order to minimize undesirable effects to neutral atoms arising from electrode magnetization.
The trap involves a pair of rods that provides a cancellation of stray electric fields to minimize excess micromotion (not shown in the figure for clarity).

The trap is driven by a 3 MHz RF field with an amplitude of 40 V.
The secular oscillation frequencies were measured to be 60 kHz in the radial direction and 12.5 kHz in the axial direction.
To load $^{40}$Ca$^{+}$ into the ion trap, an atomic beam from a Ca oven is ionized by a two-step photo-ionization process using 423 nm and 375 nm laser irradiation.
Trapped ions are Doppler cooled by 397 nm laser light, which is negatively detuned from the resonance of the $S_{1/2}-P_{1/2}$ transition and an 866 nm laser is uised for repumping from $D_{\rm{3/2}}$ for efficient laser cooling.
The photo-ionization and ion cooling lasers are arranged to propagate parallel to the ion trap axis, as depicted in Fig.\ \ref{fig:1}(b).  
During laser cooling, fluorescence is collected from the ions and focused onto a photomultiplier tube and an electron-multiplying charge coupled device (EMCCD) camera.

\begin{figure}[h]
\includegraphics[width=7cm]{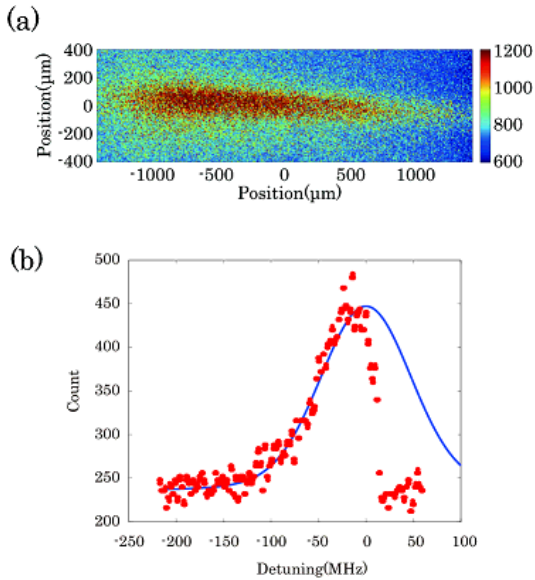}\\
\caption{\label{fig:3} (a) Image of the ion cloud. The size of the cloud is 240 $\rm{\mu}$m and 1400 $\rm{\mu}$m in the radial and the axial directions, respectively. (b) Fluorescence spectrum from the ions detected by a photomultiplier tube. The vertical and horizontal axis represent photon count and detuning of the cooling laser, respectively. The solid curve is a fitting result with a Voigt profile.} 
\end{figure}

Figure\ \ref{fig:3}(a) shows an image of the ion cloud taken by the EMCCD camera. 
The size of the ion cloud is determined by fitting the density profile with a Gaussian function, and the 1/e$^2$ widths along the radial and axial directions were 240 $\rm{\mu}$m and 1400 $\rm{\mu}$m, respectively.
Figure\ \ref{fig:3}(b) shows a fluorescence spectrum from the ion cloud obtained during Doppler cooling.
Photon counts integrated in 100 ms are plotted as a function of the frequency detuning of the cooling laser from resonance.
The solid curve is a fitted curve with a Voigt profile, which is expressed as a convolution of Lorentzian and Gaussian functions \cite{Jensen}.
The Gaussian component represents the broadening of the spectrum due to a Doppler effect, and the Lorentzian component represents the original spectral width of the $S$ - $P$ transition. The width of the Lorentzian component is fixed at 20.6 MHz (natural linewidth of the transition) and the width of the Gaussian component was a fitting parameter used to obtain the temperature of the ions from the spectrum shown in Fig.\ \ref{fig:3}(b).
Here, the data ranging from 0 to -220 MHz detuning are used for fitting.
From this analysis, the width of the Gaussian component $\Delta\nu_{\rm{G}}$ = 2$\sqrt{\rm{2}} \sigma$ = 127(6)MHz is obtained ($\sigma$ is the 1/e$^{2}$ width).
The temperature of the ion cloud and $\Delta\nu_{\rm{G}}$ are related as $T$=$\frac{m_{\rm{i}} c^{2}}{8 k_{\rm{B}} \rm{ln} \rm{2}}$($\frac{\Delta\nu_{\rm{G}}}{\rm{\nu}} ) ^{2}$ \cite{Jensen,Kielpinski}, where $m_{\rm{i}}$, $c$, $k_{\rm{B}}$, and $\nu$ are the mass of a calcium ion, the speed of light, the Boltzmann constant, and the frequency of the $S_{1/2}$ - $P_{1/2}$ transition, respectively.
From this relation, the temperature of the ions is determined to be $T$ = 2.2(2) K.
We expect the main cause for the much higher temperature than the Doppler cooling limit (490 $\rm{\mu}$K) is micromotion-induced heating.

The number of ions in the cloud is estimated from the total number of photons measured by the EMCCD camera. The number of photons emitted from each ion is calculated from the frequency detuning of the cooling laser, the photon collection efficiency of the imaging system, and the gain of the EMCCD camera.
From this estimation, we obtain an ion number $N_{\rm{i}}$ = 700(65) and an ion density in the ion cloud of $n_{\rm{i}}$ = 1.2(1) $\times$10$^{13}$ m$^{-3}$.
These values are used to calculate an elastic collision cross section between atoms and ions, as described in the next section.

By changing the loading time of the ions, we can control the number of ions in the trap and control the ion temperature from a few mK, at which the ions form a Coulomb crystal, to $\sim$K order.
In this way, we measured the elastic collisions between atoms and ions at various collision energies by changing the ion temperature.

\section {\label{sec:Results and discussion}Results and discussion}
Here, we describe the observation of elastic collisions between lithium atoms and calcium ions.
As explained in the previous section, atoms are transported so as to collide with the ion clouds in the ion trap, and we experimentally observed the loss of atoms.  
Although the temperature of the atoms is 180 $\rm{\mu K}$, the temperature of the ions is much higher. Therefore, the collision energy is determined by the kinetic energy of the ions. Since the trap depth for the atoms is much lower than the collision energy, elastic collisions are observed as the loss of atoms.

\begin{figure}[h]
\includegraphics[width=8cm]{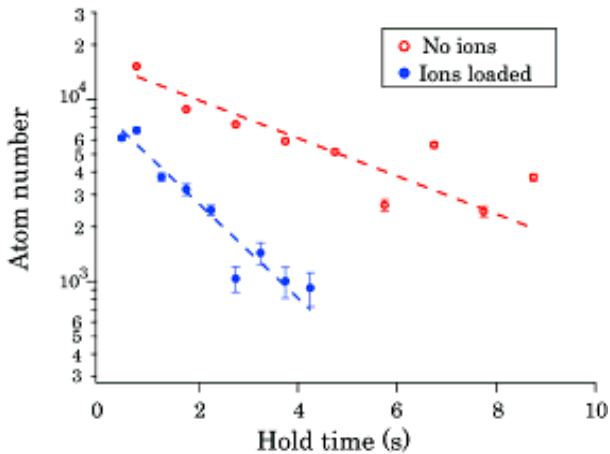}\\
\caption{\label{fig:4} Number of atoms versus atom-ion interaction time. Solid and hollow circles show data obtained when ions are loaded and unloaded in the ion trap, respectively. The dashed lines are fitting lines based on exponential decay.} 
\end{figure}

Figure\ \ref{fig:4} shows the number of atoms remaining in the ODT as a function of holding time after overlapping the atoms with the ion cloud.
The number of atoms is extracted from the axial absorption images taken immediately after release from the optical potential.
The solid and hollow circles in Fig.\ \ref{fig:4} represent the experimentally measured atomic decay with and without overlapping laser-cooled ions, respectively.
The dashed lines are fitting results with an exponential decay to the data.
A faster atom decay is seen when ions are loaded, indicating an increase of the atom loss rate due to atom-ion elastic collisions.
The atom decay rate with no ions is $\Gamma_{\rm{bg}}$ = 0.23(4) Hz, which is considered to be the loss rate due to background gas collisions.
$\Gamma_{\rm{bg}}$ + $\Gamma_{\rm{el}}$ = 0.59(7) Hz was obtained from the decay of atoms overlapped with ions, which gives an atom-ion elastic collision rate of $\Gamma_{\rm{el}}$ = 0.36(8) Hz.

Atom loss by elastic collision can be expressed as $\dot {N_{\rm{a}}}$ = $\int \dot{n}_{\rm{a}} d \textbf{r}$ = -$\int n_{\rm{i}} \sigma_{\rm{els}} v_{\rm{rel}} n_{\rm{a}} d\textbf{r}$ = - $\Gamma_{\rm{els}} N_{\rm{a}}$ ($\Gamma_{\rm{els}}$ = $n_{\rm{i}}\sigma_{\rm{els}}v_{\rm{rel}}$).
Here, we assume the ion density is homogeneous.
$N_{\rm{a}}$, $n_{\rm{a}}$, $\sigma_{\rm{els}}$, $v_{\rm{rel}}$, and $n_{\rm{i}}$  are the number of atoms, atom density, the elastic scattering cross section, the relative velocity of atoms and ions, and the ion density.
By substituting $n_{\rm{i}}$ and $v_{\rm{rel}} = \sqrt{\frac{3 k_{\rm{B}}T}{\mu}}$ ($T \approx T_{\rm{ion}}$) with $\rm{\mu}$, the reduced mass of a $^{\rm{6}}$Li and a $^{\rm{40}}$Ca$^{\rm{+}}$ ion, $\sigma_{\rm{els}}$ = 3.3(9) $\times$10$^{-16}$ $\rm{m}^{2}$ is obtained.

On the other hand, the atom-ion scattering cross section is expressed within a certain collision energy range using a semi-classical model \cite{Cote} as,

\begin{eqnarray}
\label{eq:1}
\sigma_{\rm{els}} = \pi \left( 1 + \frac{\pi^{2}}{16}\right) \left( \frac{\mu C_{4}^{2}}{\hbar^{2}} \right)^{1/3} E^{-1/3}
\end{eqnarray}

Here, $\hbar$ and $E \approx E_{\rm{ion}} = k_{\rm{B}} T_{\rm{ion}}$ are Planck's constant and the collision energy, respectively, and $C_{4} = - \frac{\check{\alpha}q^{2}}{4\pi\epsilon_{0}}$.
$q$, $\epsilon_{\rm{0}}$, and $\check{\alpha}$ are the electron charge, the electron permittivity in vacuum, and the DC polarizability of lithium, respectively.
For the lithium atom, $\check{\alpha}$ = 164 $a_{\rm{0}}^{3}$ \cite{Teachout}, where $a_{\rm{0}}$ is the Bohr radius.
By substituting $E = k_{\rm{B}}T_{\rm{ion}}$ ($T_{\rm{ion}}=2.2$ K) into Eq.\ (\ref{eq:1}), $\sigma_{\rm{cal}}$ = 4.7 $\times$10$^{-16}$ m$^{2}$ results, and this is consistent with the experimental results.

\begin{figure}[h]
\includegraphics[width=8cm]{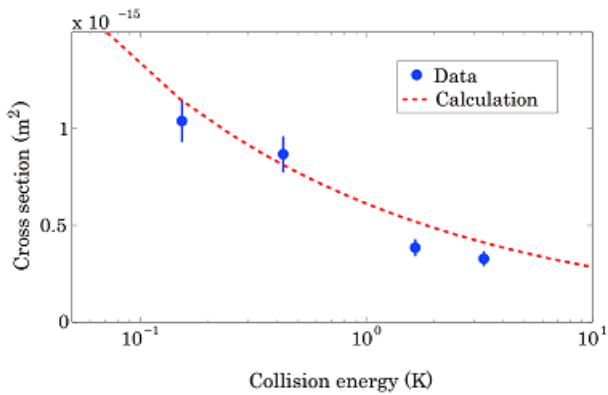}\\
\caption{\label{fig:5} Elastic collision cross section vs. collision energy. The dashed curve is a theoretical prediction based on a semi-classical collision model.} 
\end{figure}

We also measured the elastic scattering cross sections at various collision energies.
The results are summarized in Fig.\ \ref{fig:5}.
To change the collision energy, i.e. the ion temperature, the number of ions in the ion trap was varied.
The solid circles are the data with error bars, and typical atom number fluctuations (10$\%$) were taken into account.  
The dashed curve is a calculated result from Eq.\ (\ref{eq:1}).
As shown in Fig.\ \ref{fig:5}, the theoretical prediction well-reproduced the experimental data, indicating that the semi-classical theory is valid for collision energies ranging from 100 mK to 3 K. 

\section {\label{sec:Conclusion} Conclusion and outlook}
We have constructed an experimental setup to create an ultracold atom-ion mixture, and observed the elastic collisions between fermionic lithium atoms and calcium ions at the energy range from 100 mK to 3K.
In this experiment, atoms were trapped by optical means and were successively transported to the position of an ion cloud using an optical tweezer technique. We observed elastic collisions as the loss of atoms from the trap. The collision-energy dependence of the elastic scattering cross section was experimentally observed to be consistent with a semi-classical collision model.

This work is an essential step toward creating atom-ion mixtures in the ultracold regime, where quantum scattering dominates the dynamics of the system.
The threshold collision energy of the $s$-wave scattering between atoms and ions is given as $E_{s} = \frac{\hbar^{4}}{2 \mu ^{2} C_{4}}$, and $E_{\rm{s}}$/$k_{\rm{B}}$ = 10 $\rm{\mu K}$ is calculated for the $^{6}$Li - $^{40}$Ca$^{+}$ combination.
10 $\rm{\mu}$K is expected to be achievable by a precise micromotion cancellation and the preparation of a motional  ground state of ions by a resolved-sideband cooling technique \cite{Monroe}.
Furthermore, the heating rate due to micromotion induced by atom-ion interaction is estimated to be 0.5 $\rm{\mu}$K for the $^{6}$Li - $^{40}$Ca$^{+}$ combination. 
Therefore, our combination consisting of light neutral atoms is suitable for the realization of ultracold atom-ion interactions and ultracold chemistry.
The next challenge is to observe atom-ion collisions in a quantum mechanical regime using ultracold ions and atoms at $\rm{\mu}$K temperatures.

We would like to thank Kensuke Matsubara, Kunihiro Okada, Nobuyasu Shiga and Kazuhiro Hayasaka for their advice on the ion trap.
We thank Ken'ichi Nakagawa for stimulating discussion.

%\end{multicols}
\end{document}